\documentclass[prl,aps,showpacs,preprint,noshowpacs]{revtex4}

\usepackage{epsfig}

\begin{document}

\title{Magnetic Bubblecade Memory}
\author{Kyoung-Woong Moon$^{1,\dag}$, Duck-Ho Kim$^{2,\dag}$, Sang-Cheol Yoo$^{2,3}$, Soong-Geun Je$^{2}$, Byong Sun Chun$^{1}$, Wondong Kim$^{1}$, Byoung-Chul Min$^{3}$, Chanyong Hwang$^{1,*}$, and Sung-Bong Choe$^{2,*}$}
\affiliation{$^1$Center for Nanometrology, Korea Research Institute of Standards and Science, Daejeon 305-340, Republic of Korea}
\affiliation{$^2$CSO and Department of Physics, Seoul National University, Seoul 151-742, Republic of Korea}
\affiliation{$^3$Spin Convergence Research Center, Korea Institute of Science and Technology, Seoul 136-791, Republic of Korea}

\begin{abstract}
Unidirectional motion of magnetic domain walls is the key concept underlying next-generation domain-wall-mediated memory and logic devices. 
Such motion has been achieved either by injecting large electric currents into nanowires or by employing domain-wall tension induced by sophisticated structural modulation. 
Herein, we demonstrate a new scheme without any current injection or structural modulation. 
This scheme utilizes the recently discovered chiral domain walls, which exhibit asymmetry in their speed with respect to magnetic fields. 
Because of this asymmetry, an alternating magnetic field results in the coherent motion of the domain walls in one direction. 
Such coherent unidirectional motion is achieved even for an array of magnetic bubble domains, enabling the design of a new device prototype---magnetic bubblecade memory---with two-dimensional data-storage capability.
\end{abstract}

\pacs{75.78.-n, 75.70.Kw, 85.70.Li, 85.70.-w}

\maketitle

Recent progress in the control of magnetic domain walls (DWs) has suggested a number of prospective opportunities for next-generation DW-mediated devices~\cite{PK,AD,NOT,HS,KJ}.
Among these, coherent unidirectional DW motion has been proposed to replace the mechanical motion of magnetic media in hard-disk drives, thereby enabling the creation of a solid-state nonvolatile data-storage device---so-called racetrack memory---with high storage capacity, low power, and high mechanical stability~\cite{PK}.
Such coherent unidirectional motion was first achieved by injecting current into magnetic nanowires~\cite{HS,KJ,TV}. 
In this scheme, current-induced spin-transfer~\cite{TV} and spin-orbit torques~\cite{JK,YF,KS} exert forces on DWs by transferring electron spins to the local magnetic moment, resulting in DW motion along the direction of force. 
It is therefore possible to realize the unidirectional motion of multiple DWs~\cite{HS,KJ}, leading to the recent development of DW shift registers. 
Such DW motion, however, requires a high threshold current, which is inevitably accompanied by high Joule heating that may cause severe artifacts~\cite{CY}. 
Structural modulation of the nanowires has therefore been proposed to reduce the threshold current by introducing DW tension~\cite{KW}. 
With wedge-shaped modulation, the DW tension exerts a force on the DWs to reduce the tension energy and consequently, facilitates the DW motion toward the apex edge~\cite{AH}. 
It has been revealed that periodic structural modulation allows DW-tension-induced unidirectional motion to be solely driven by the magnetic field without any current injection, as demonstrated by magnetic-ratchet shift registers~\cite{FK}. 
It has also been demonstrated that vertical composition modulation leads to unidirectional DW motion along the vertical direction~\cite{RL}.    
These schemes are, however, extremely sensitive to tiny structural irregularities in the devices and thus, require highly sophisticated nanofabrication processes.
Here, we demonstrate a new scheme for unidirectional DW motion based on skyrmion-like magnetic bubble domains.
This scheme is applicable even to unpatterned films in the absence of any current injection or structural modulation.

The magnetic skyrmion is a topological object in which the internal spins whirl around the core in all directions and thus, shield the core spins from outer spins of the opposite orientation~\cite{NG,KK,UK,JH,NS,SP,JL}.
Magnetic skyrmions have been observed in several helical magnets, where the helical spin alignment is caused by the Dzyaloshinskii-Moriya interaction (DMI)~\cite{DZ,MR}.
Recently, it has been observed that metallic ferromagnetic multilayer films also exhibit finite DMI because of their asymmetric layer structure, resulting in skyrmion-like magnetic bubble domains with a N{\' e}el DW configuration~\cite{GC1,GC2}.

\begin{figure}
\includegraphics[width=16.8 cm]{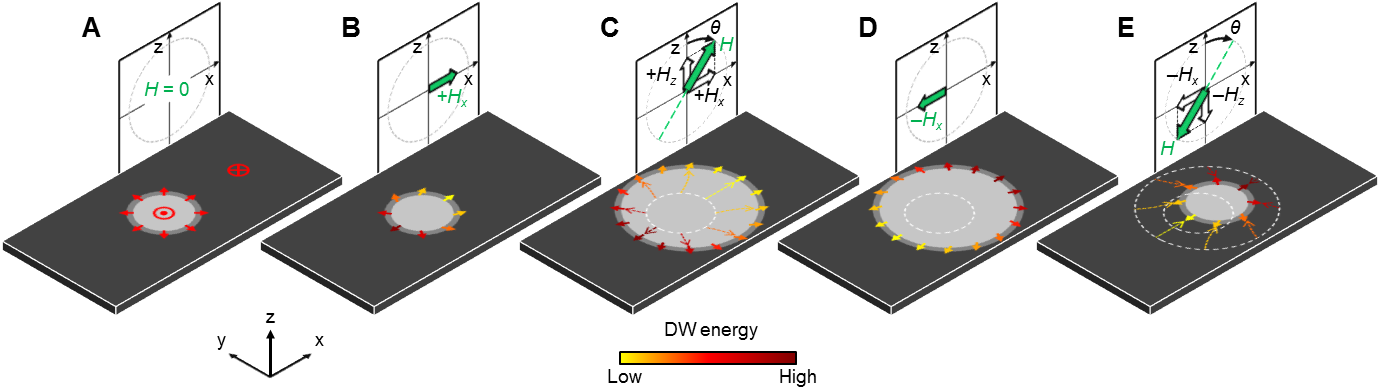}
\caption{
({\bf A}) Illustration of a bubble domain (bright circle) and the DW (grey ring), surrounded by a domain of opposite magnetization (dark area).
The red symbols and arrows indicate the direction of the magnetization inside the DW and domains.
({\bf B}) Asymmetric DW-energy distribution under an in-plane magnetic field $+H_{x}$ (green arrow), as visualized by the color contrast of the arrows on the DW according to the scale bar at the bottom.
({\bf C}) Asymmetric DW expansion under a magnetic field $H$ (=($+H_{x}$, $+H_{z}$)) with a tilting angle ${\theta}$.
({\bf D}) Asymmetric DW-energy distribution under the reversed in-plane magnetic field $-H_{x}$.
({\bf E}) Asymmetric DW shrinkage under the reversed magnetic field ($-H_{x}$, $-H_{z}$).
The dashed circles represent the previous DW positions.}
\label{Fig1}
\end{figure}

We demonstrate that a sequence of applying magnetic fields leads to a unidirectional motion of magnetic bubble domains.
Figure~\ref{Fig1}A illustrates a skyrmion-like bubble domain with a N{\' e}el DW configuration, where the magnetization $\hat{m}_{\rm DW}$  (red arrows) inside the DW is oriented radially outward in all directions.
This bubble expands or shrinks circularly under an out-of-plane magnetic field $H_{z}$, because of the rotational symmetry with respect to the center of the bubble.
If one applies an in-plane magnetic field $H_{x}$, then the rotational symmetry is broken by the Zeeman interaction between $H_{x}$ and $\hat{m}_{\rm DW}$ , resulting in an asymmetric distribution of the DW energy, as indicated by the colored arrows in Fig.~\ref{Fig1}B.
This bubble domain then exhibits asymmetric expansion under $H_{z}$ (Fig.~\ref{Fig1}C) because the DW speed depends on the DW energy~\cite{SG,HB,TV2}.
At this instant, if one reverses the polarity of the in-plane magnetic field (i.e., applies $-H_{x}$), then the asymmetry in the DW-energy distribution is also reversed (Fig.~\ref{Fig1}D).
With applying $H_{z}$, this bubble domain shrinks toward a different location from the original position of the domain (Fig.~\ref{Fig1}E).
Consequently, the center of the bubble shifts along the $x$ axis from the original position.
Such a shift of the center can be continuously generated along the same direction by repeating the process illustrated in Figs.~\ref{Fig1}C and E, in which collinear magnetic fields ($+H_{x}$, $+H_{z}$) and ($-H_{x}$, $-H_{z}$) are alternately applied.
Therefore, unidirectional bubble motion can be achieved by applying an alternating magnetic field generated by a single coil that is tilted by an angle ${\theta}$ ($={\rm atan}⁡(H_{x}/H_{z})$) to the film normal.

\begin{figure}
\includegraphics[width=16.8 cm]{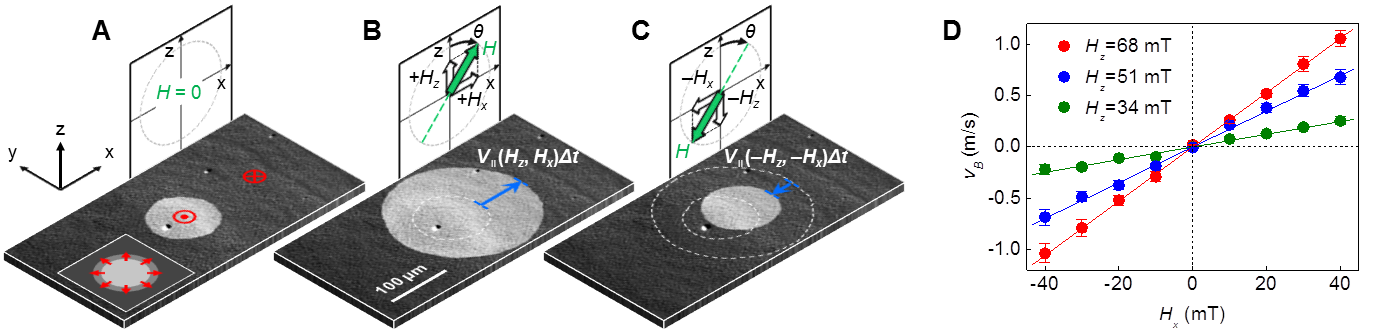}
\caption{
Experimental verification of the unidirectional bubble motion from Pt/Co/Pt film.
({\bf A}) MOKE image of the initial bubble domain (bright circle) surrounded by a domain of opposite magnetization (dark area).
The red symbols indicate the direction of the magnetization in the domains.
The inset illustrates the expected $\hat{m}_{\rm DW}$  (red arrows) in the right-handed chiral DW configuration.
({\bf B}) The expanded bubble domain after application of a ($+H_{x}$,$+H_{z}$) pulse ($H_{x}$= 30 mT, $H_{z}$= 4 mT, $\Delta t$= 100 ms), where $\Delta t$ is the pulse duration time.
The blue arrow indicates the DW displacement $V_{\parallel}(H_{z},H_{x})\Delta t$. The dashed circle represents the initial DW position.
({\bf C}) The shrunken bubble domain after application of a ($-H_{x}$,$-H_{z}$) pulse.
The blue arrow indicates the DW displacement $V_{\parallel}(-H_{z},-H_{x})\Delta t$.
The dashed circles represent the previous DW positions.
({\bf D}) Measured $V_{B}$ with respect to $H_{x}$ for several $H_{z}$.}
\label{Fig2}
\end{figure}

The predicted behavior discussed above can be readily verified for Pt/Co/Pt films.
Recent studies have revealed that these films have a positive DMI and thus exhibit the right-handed chiral DW configuration.
Figure~\ref{Fig2}A presents an image of a bubble domain captured using a magneto-optical Kerr effect (MOKE) microscope~\cite{KW}.
Because of the right-handed chirality, $\hat{m}_{\rm DW}$  is expected to be oriented outward, as illustrated in the inset.
By applying an alternating magnetic field to this bubble domain, unidirectional bubble motion was successfully accomplished, as seen in Figs.~\ref{Fig2}A-C.
The exact conformity of these images with Fig.~\ref{Fig1} proves the principle of the present scheme.

The speed $V_{B}$ of the bubble motion follows the average rate of DW motion under the alternating magnetic-field pulses.
The forward and backward motions of the DW (blue arrows in Figs.~\ref{Fig2}B and C) yield the relation $V_{B}$=$[V_{\parallel}(H_{z},H_{x})+V_{\parallel}(-H_{z},-H_{x})]/2$, where $V_{\parallel}$ is the DW speed at the rightmost point of the bubble~\cite{Supp}.
The measured $V_{B}$ is plotted with respect to $H_{x}$ for several $H_{z}$ (Fig.~\ref{Fig2}D).
This plot clearly demonstrates that $V_{B}$ is proportional to $H_{x}$ within the experimental range of $H_{x}$, yielding the expression $V_{B}\cong\rho_{1}(H_{z})H_{x}$.
According to Ref.~\cite{Supp}, the coefficient $\rho_{1}(H_{z})$ is given by $\{C_{1}{\rm ln}⁡[V_{0}/|V_{\parallel}(H_{z},0)|]\}V_{\parallel}(H_{z},0)$ in the DW creep regime~\cite{MT,LM}, where $C_{1}$ is a constant related to the Zeeman contribution to the DW energy and $V_{0}$ is the characteristic speed.
In the present sample, ${C_{1}{\rm ln}⁡[V_{0}/|V_{\parallel}(H_{z},0)|]}$ is estimated to be approximately (86 mT)$^{-1}$ for the maximum $H_{z}$ (68 mT) from the present coil and thus, a  $V_{B}$ is achieved up to approximately 46{\%} of $V_{\parallel}(H_{z},0)$ under the maximum $H_{x}$ (40 mT). Optimizing the design of the coil will further enhance $V_{B}$, as the present maximum $V_{B}$ ($\sim$1 m/s) is not limited by the sample.

\begin{figure}
\includegraphics[width=16.8 cm]{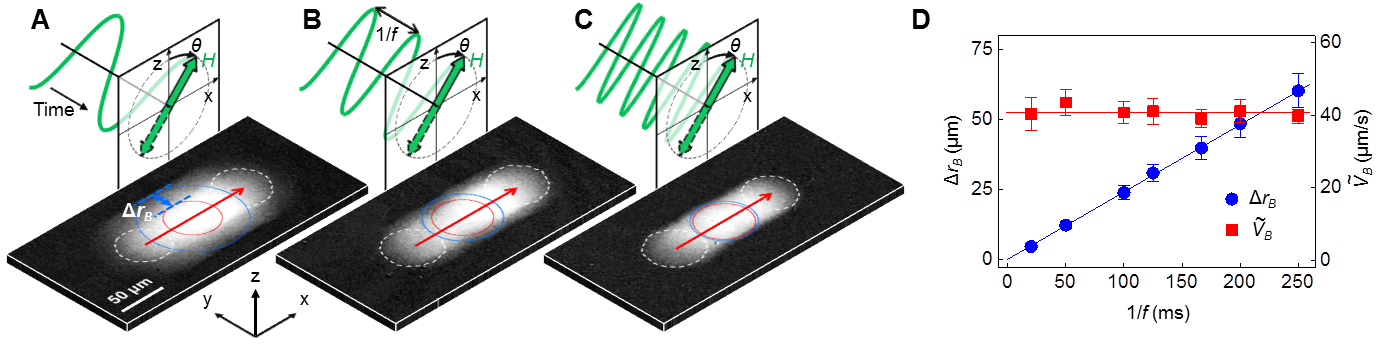}
\caption{
Accumulated MOKE images acquired during bubble motion for various $f$ values, 10 Hz ({\bf A}), 20 Hz ({\bf B}), and 50 Hz ({\bf C}), respectively.
Alternating sinusoidal magnetic field was applied with an amplitude 3.7 mT and a tilting angle $34^{\circ}$, and the accumulation time was 3 s.
The red (blue) circle represents the smallest (largest) bubble observed during the motion.
The blue arrows indicate $\Delta r_{B}$ between the largest and smallest bubbles.
The red arrows indicate the displacement of the bubble center between the initial and final positions (white dashed circles) over the accumulation time.
({\bf D}) Measured $\Delta r_{B}$ and $\tilde{V}_{B}$ values with respect to $f$.
}
\label{Fig3}
\end{figure}

The variation $\Delta r_{B}$ in the radius of the bubble during its motion can be controlled by adjusting the frequency $f$ of the alternating magnetic field.
Figures~\ref{Fig3}A-C present images of the bubble motion driven by alternating sinusoidal magnetic field with $f$= 10 Hz (A), 20 Hz (B), and 50 Hz (C), respectively.
Note that each image was accumulated over a period of 3 s during bubble motion and thus, the length (red arrow) of the gray area represents the bubble displacement during the image-accumulation time.
Additionally, the width (blue arrow) of the light-gray boundary represents $\Delta r_{B}$ between the smallest (red circle) and largest (blue circle) bubbles.
Figure~\ref{Fig3}D provides a plot of the measured $\Delta r_{B}$ and the average bubble speed $\tilde{V}_{B}$ values with respect to $f$.
The figure clearly demonstrates that $\Delta r_{B}$ is inversely proportional to $f$. Because $\tilde{V}_{B}$ remains nearly unchanged irrespective of $f$, one can independently reduce $\Delta r_{B}$ down to the limiting value defined by the bandwidth of the coil without changing $\tilde{V}_{B}$.

\begin{figure}
\includegraphics[width=5.5 cm]{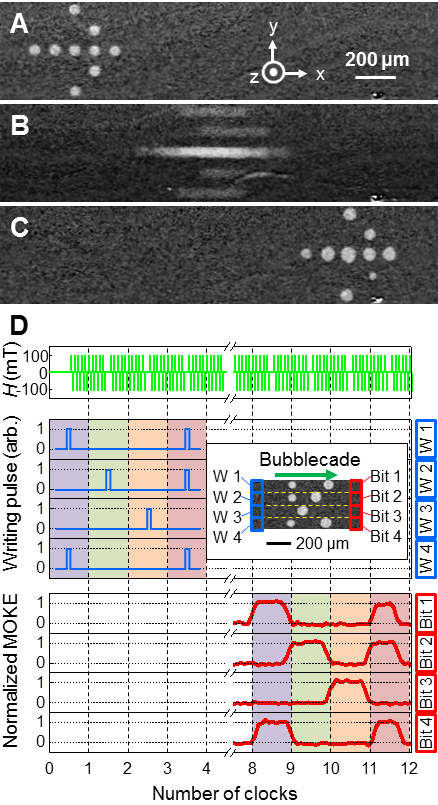}
\caption{
Experimental demonstration of\, \textquoteleft bubblecade,\textquoteright \, the two-dimensional coherent unidirectional motion of multiple bubbles.
({\bf A}) MOKE image of an arbitrary 5$\times$5 array of bubbles, initially written on the leftmost side of the film.
({\bf B}) Accumulated MOKE image acquired during bubble motion under alternating magnetic-field pulses ($H$= 106 mT, $\theta$= $71^{\circ}$, $\Delta t$= 20 ms).
({\bf C}) MOKE image of the bubble array after displacement to the rightmost side of the film.
({\bf D}) The operation timetable of 4-bit magnetic bubblecade memory.
The top panel presents the alternating magnetic-field pulses ($H$= 106 mT, $\theta$= $71^{\circ}$, $\Delta t$= 20 ms) that act as the operation clock.
The next four panels show the pulses for the thermomagnetic writing of the bubbles on each bit in the writing section.
The last four panels show the signals from the CCD pixels corresponding to each bit of the reading section.
The inset presents the device structure, with the writing (blue box) and reading (red box) sections indicated.
}
\label{Fig4}
\end{figure}

Finally, the present scheme was applied to a two-dimensional bubble array. For this purpose, an arbitrary 5$\times$5 array pattern of bubbles (Fig.~\ref{Fig4}A) was initially created on the film using the thermomagnetic writing method~\cite{Supp}.
Under the application of alternating magnetic pulses, all bubbles exhibited coherent unidirectional motion, as shown by the image (Fig.~\ref{Fig4}B and movie S1) captured during the pulses.
Exactly the same bubble-array pattern was maintained even after traveling over 2 mm (Fig.~\ref{Fig4}C).
Therefore, the observed two-dimensional coherent unidirectional motion of the bubbles---hereafter referred to as the bubblecade---can be used to replace the mechanical motion of the magnetic media with respect to read and write sensors, enabling a new device prototype \,\textquoteleft magnetic bubblecade memory.\textquoteright

The writing and reading operation schemes of bubblecade memory are also demonstrated. Figure~\ref{Fig4}D illustrates the operation timetable for a 4-bit magnetic bubblecade memory.
The top panel shows the alternating magnetic pulses that act as the operation clock.
The next four panels specify the bubble-writing pulses, which are applied to each bit of the writing section (blue box) of the device depicted in the inset.
At present, bubble writing is achieved using the thermomagnetic writing scheme, but it may also be possible to implement using the spin-transfer torque writing scheme with a nanopillar structure~\cite{SP}.
The bottom four panels illustrate the reading signal from each bit of the reading section (red box) of the device.
At present, the reading signals are detected by the corresponding pixels of a charge-coupled device (CCD) camera,
but it may also be possible to read out these signals using tunneling magnetoresistive sensors in the future~\cite{MD}.
The figure clearly shows that all written two-dimensional data bits are successively retrieved from the reading section, demonstrating shift-register-based memory operation (movie S2).

In summary, we present here a proof-of-principle experiment demonstrating the two-dimensional coherent unidirectional motion of multiple bubble domains, of which the speed is a significant fraction of the DW speed.
Such bubble motion is attributed to the helical magnetic configuration caused by the asymmetric layer structure and therefore, further exploration of materials and layer combinations~\cite{DH} together with the optimization of the coil design will further enhance the potential of this technology for various applications.
The present scheme can eliminate the necessity for the mechanical motion of the media in hard-disk drives and thus, enables the development of a new prototype solid-state data-storage device, the so-called magnetic bubblecade memory.

\

\
This work was supported by Ministry of Science, ICT and Future Planning through Center for Advanced Meta-Materials and also supported by the National Research Foundation of Korea (NRF) grant funded by the Korea government (MSIP) (2012-003418, 2008-0061906).
S.-C.Y. and B.C.M. were supported by the KIST institutional program and the Pioneer Research Center Program of MSIP/NRF (2011-0027905).

\newpage
\makeatletter
\@fpsep\textheight
\makeatother

\newpage

\newpage

\end{document}


\title{{\Large Supplementary Materials for} 

\vspace{5 mm}

Magnetic Bubblecade Memory}
\author{Kyoung-Woong Moon, Duck-Ho Kim, Sang-Cheol Yoo, Soong-Geun Je, Byong Sun Chun, Wondong Kim, Byoung-Chul Min, Chanyong Hwang, and Sung-Bong Choe}

\pacs{75.78.-n, 75.70.Kw, 85.70.Li, 85.70.-w}

\maketitle
\vspace{5 mm}
{\setlength\parindent{0pt} \large \bf This PDF file includes:}
\vspace{2 mm}

    Materials and Methods

    Supplementary Text 

    Fig. S1 

    References and Notes

\vspace{3 mm}

{\setlength\parindent{0pt}\large \bf Other Supplementary Material for this manuscript includes the following:}
\vspace{2 mm}

Movies S1 and S2
\vspace{10 mm}

{\setlength\parindent{0pt}\bf Materials and Methods}
\vspace{2 mm}

{\setlength\parindent{0pt}\underline {Sample preparation}}
\vspace{2 mm}

For this study, metallic ferromagnetic Ta/Pt/Co/Pt films were deposited on Si substrates with 100-nm-thick SiO$_{2}$ layer by means of the dc-magnetron sputtering.
The thicknesses of the Ta and Co layers are fixed to 5.0 and 0.3 nm, respectively, and the thicknesses of the upper and lower Pt layers are adjusted from 1.0 to 3.0 nm to tune the magnetic properties~\cite{DH}.
To enhance the sharpness of the layer interfaces, the films were deposited with a small deposition rate (0.25 \AA/sec) through adjustment of the Ar sputtering pressure ($\sim$2 mTorr) and power ($\sim$10 W).
All the films exhibit clear circular domain expansion with weak pinning strength.
The results in Figs. 2 and 3 were obtained from 5.0-nm Ta/2.5-nm Pt/0.3-nm Co/3.0-nm Pt film that shows the fastest bubble speed under the present experimental condition, possibly due to the weak coercive field (7.1 mT).
The results in Fig. 4 were obtained from 5.0-nm Ta/2.5-nm Pt/0.3-nm Co/1.0-nm Pt film that allows regular bubble-array writing with small irregularities due to the relatively large coercive field (16.2 mT).

\vspace{5 mm}
{\setlength\parindent{0pt}\underline {Experimental setup and procedure}}
\vspace{2 mm}

The magnetic domain images were observed by use of a magneto-optical Kerr effect (MOKE) microscope equipped with a charge-coupled device (CCD) camera on the focal plane~\cite{KW}.
To apply the magnetic field onto the films, two electromagnets and two small coils are attached to the sample stage.
One of the electromagnets is used to apply the in-plane magnetic field bias up to 200 mT.
The smallest coil ($\sim$1 mm in radius) is used to apply the out-of-plane magnetic field pulses up to 68 mT with a fast rising time ($<1$ $\mu$s).
The combination of the in-plane electromagnet and the smallest coil was used to obtain the results shown in Fig. 2. The other coil ($\sim$2 mm in radius) is designed to apply the alternating sinusoidal magnetic field with adjustable tilting angle, which was used to obtain the results shown in Fig. 3. For field uniformity over the wide range ($>2$ mm) of the film, two electromagnets were used to apply the in-plane and out-of-plane magnetic fields to obtain the results shown in Fig. 4.

To write bubble domains, the magnetization of the film is first saturated by applying an out-of-plane magnetic field pulse (–30 mT, 1 s).
A laser beam (60 mW) is then focused on a small spot (${\sim}1$ $\mu$m in diameter) of the film, causing reduction of the coercive field inside the spot by increasing the temperature.
At this instant, a reversed magnetic field pulse (8 mT, 12 ms) is applied.
Since the strength of the reversed magnetic field is adjusted to be slightly larger than the reduced coercive field inside the spot, the magnetization reversal occurs only in the area of the spot.
Consequently, a bubble-shaped reversed domain is created.
By repeating this procedure, array of the bubble domains is created.

\vspace{5 mm}
{\setlength\parindent{0pt}\underline {Movie S1}}
\vspace{2 mm}

The present movie was recorded during the bubbles motion.
Each image was taken after application of alternating magnetic field pulses ($H=\pm$106 mT, $\theta=71^{\circ}$, $\Delta t= 20$ ms). The movie shows the images with 20 frames/s.

\vspace{5 mm}
{\setlength\parindent{0pt}\underline {Movie S2}}
\vspace{2 mm}

The present movie was recorded during the magnetic bubblecade memory operation as shown by Fig. 4D.
During a clock for one-bit shift, 11 frames of images were taken: 10 frames were taken after each application of a pair of alternating magnetic field pulses ($H=\pm$106 mT, $\theta=71^{\circ}$, $\Delta t= 20$ ms) and 1 frame was taken for stationary image.
The movie shows the images with 20 frames/s.

\vspace{10 mm}
{\setlength\parindent{0pt}\bf Supplementary Text}
\vspace{2 mm}

{\setlength\parindent{0pt}\underline {Speed of the bubblecade motion}}
\vspace{2 mm}

By adopting the Taylor expansion with respect to $H_{x}$, the DW speed $V_{\parallel}$ at the rightmost point of the bubble domain can be written as
\begin{equation}
V_{\parallel}(H_{z},H_{x})=V_{\parallel}(H_{z},0)+{\sum}_{n=1}^{\infty}\rho_{n}(H_{z})H_{x}^{n},
\tag{S1}
\label{S1}
\end{equation}
where $\rho_{n}(H_{z})\equiv{\frac{1}{n!}\frac{{\partial}^{n}V_{\parallel}(H_{z},H_{x})}{\partial H_x^n}\Big|_{H_{x}=0}}$.
Since $V_{\parallel}$ is an odd function with respect to $H_z$, $V_{\parallel}$($-H_z$,$-H_x$) is equal to $-V_{\parallel}$($H_z$,$-H_x$) i.e.
\begin{equation}
V_{\parallel}(-H_{z},-H_{x})=-V_{\parallel}(H_{z},0)-{\sum}_{n=1}^{\infty}\rho_{n}(H_{z})(-H_{x})^{n}.
\tag{S2}
\label{S2}
\end{equation}
From the relation $V_B$=[$V_{\parallel}$($H_z$,$H_x$)+$V_{\parallel}$($-H_z$,$-H_x$)]/2, the speed $V_B$ of the bubble motion can be thus expressed as
\begin{equation}
V_{B}(-H_{z},-H_{x})={\sum}_{n=0}^{\infty}\rho_{2n+1}(H_z)H_{x}^{2n+1},
\tag{S3}
\label{S3}
\end{equation}
The experimental observation (Fig. 2D) indicates that it is good enough to approximate Eq. (S3) as
\begin{equation}
V_{B}(H_{z},H_{x})\cong{\rho}_{1}(H_z)H_{x},
\tag{S4}
\label{S4}
\end{equation}
within the present experimental range of $H_x$, by confirming that the higher-order terms are negligible compared to the linear term.

According to Ref.~\cite{SG}, the DW energy density $\sigma_{\rm DW}$  is given by a function of $H_x$ as
\[
\sigma_{\rm DW} = \left\{
                               \begin{array}{ll}
                                \sigma_{0}-\frac{\pi \lambda M_{\rm S}}{2 H_{\rm D}}(H_x +H_{\rm DMI})^{2}  &{\rm for\,} |H_x +H_{\rm DMI}|<H_{\rm D}\\
                                \sigma_{0}+2 K_{\rm D}\lambda-\pi \lambda M_{\rm S}|H_{x}+H_{\rm DMI}|     &{\rm otherwise}
                               \end{array}
\right. ,
\tag{S5}
\]
where $\sigma_{0}$  is the DW energy of the Bloch configuration, $\lambda$ is the DW width, $M_{\rm S}$  is the saturation magnetization, and $H_{\rm DMI}$  is the DMI-induced effective magnetic field.
Here, $H_{\rm D}(\equiv4 K_{\rm D}/\pi M_{S}$) is the DW anisotropy field that is required to rotate $\hat{m}_{\rm DW}$  from the Bloch configuration to the N\'{e}el configuration, where $K_{\rm D}$  is the DW anisotropy constant.
Based on the assumption that the dependence of $V_{\parallel}$ on $H_{x}$ is solely attributed to the variation of $\sigma_{\rm DW}$  due to $H_x$, one finds the relation
\begin{equation}
\rho_{1}\equiv \frac{\partial V_{\parallel}}{\partial H_{x}}\Big|_{H_{x}=0}=\frac{\partial V_{\parallel}}{\partial \sigma_{\rm DW}}\Big|_{\sigma_{\rm DW}=\sigma_{\rm DW}(0)}\cdot\frac{d \sigma_{\rm DW}}{d H_x}\Big|_{H_x =0},
\tag{S6}
\label{S6}
\end{equation}
which is then written as
\[
\rho_{1} = \left\{
                \begin{array}{ll}
                -\pi\lambda M_{\rm S} \frac{H_{\rm DMI}}{H_{\rm D}}\frac{\partial V_{\parallel}}{\partial \sigma_{\rm DW}}\Big|_{\sigma_{\rm DW}=\sigma_{\rm DW}(0)}    &{\rm for\,} |H_{\rm DMI}|<H_{\rm D}\\
                -\pi\lambda M_{\rm S} {\rm sgn}(H_{\rm DMI}) \frac{\partial V_{\parallel}}{\partial \sigma_{\rm DW}}\Big|_{\sigma_{\rm DW}=\sigma_{\rm DW}(0)}    &{\rm otherwise}
                \end{array}
\right. ,
\tag{S7}
\]
where ${\rm sgn}(H_{\rm DMI})$ denotes the sign of $H_{\rm DMI}$.

In the creep regime, $V_{\parallel}$ follows the creep scaling law $V_{\parallel}(H_{z},H_{x})$=$V_{0}{\rm exp}⁡[-\alpha(H_{x}) H_{z}^{-1/4}]$, where $V_0$ is a characteristic speed and $\alpha$ is a constant related to the scaling energy constant, the critical magnetic field, and the thermal fluctuation energy~\cite{KJ2}. By use of the relation $\alpha(H_{x})\propto[\sigma_{\rm DW}(H_{x})]^{1/4}$ proposed in Refs.~\cite{SG,HB}, one finds the relation
\begin{equation}
\frac{\partial V_{\parallel}}{\partial \sigma_{\rm DW}}\Big|_{\sigma_{\rm DW}=\sigma_{\rm DW}(0)}=\frac{1}{4\sigma_{\rm DW}(0)}V_{\parallel}(H_{z},0){\rm ln}\Big( \frac {V_0}{|V_{\parallel}(H_{z},0)|}\Big).
\tag{S8}
\label{S8}
\end{equation}
Since $|H_{\rm DMI}|>H_{\rm D}$  in the present Pt/Co/Pt films as demonstrated in Ref.~\cite{SG} and $H_{\rm DMI}>0$ in the present experimental condition, $V_B$ can be finally written as
\begin{equation}
V_{B}(H_{z},H_{x})\cong{C}_{1}{\rm ln}\Big( \frac {V_0}{|V_{\parallel}(H_{z},0)|}\Big)V_{\parallel}(H_{z},0)H_{x},
\tag{S9}
\label{S9}
\end{equation}
where $C_{1}\equiv\frac {\pi\lambda M_{\rm S}} {4 \sigma_{\rm DW}(0)} {\rm sgn}(H_{\rm DMI})$.

For alternating sinusoidal magnetic field ($H_{x} {\rm sin}\omega t,H_{z}{\rm sin}\omega t$), the average speed $\tilde{V}_B$ of the bubble motion can be written by
\begin{equation} 
\begin{split}
\tilde{V}_B (H_{z},H_{x})& \cong \frac{C_1 \omega}{\pi}\int_{0}^{\pi / \omega} {\rm ln} \Big( \frac {V_0}{|V_{\parallel}(H_{z}{\rm sin}\omega t,0)|}\Big)V_{\parallel}(H_{z}{\rm sin}\omega t,0)H_{x}{\rm sin}\omega t dt  \\
 & \cong C_2 V_B (H_{z},H_{x})
\end{split},
\tag{S10}
\label{S10}
\end{equation}
where $C_2$=$\frac{1}{\pi}\int_{0}^{\pi}{\rm exp} \Big[ {\rm ln} \Big( \frac {V_{0}}{|V_{\parallel}(H_{z},0)|} \Big) \Big( 1-({\rm sin}\tau )^{\frac{1}{4}}\Big) \Big] ({\rm sin}\tau)^{\frac{3}{4}}d\tau  $.
Numerical evaluation reveals that $C_2$ is a slowly-varying function of $V_{\parallel}(H_{z},0)$ within the range from 0.3 (for $V_{\parallel}(H_{z},0)$$\cong$ 1 mm/s) to 0.5 (for $V_{\parallel}(H_{z},0)$$\cong$ 100 m/s) in the present samples.

\vspace{5 mm}
{\setlength\parindent{0pt}\underline {Direction of the bubblecade motion}}
\vspace{2 mm}

According to Eq. (S9), the direction of the bubble motion i.e. ${\rm sgn}(V_{B})$ is given as
\begin{equation}
{\rm sgn}(V_{B})={\rm sgn}(H_{\rm DMI} V_{\parallel}(H_{z},0)H_{x}).
\tag{S11}
\label{S11}
\end{equation}
Since the direction of $V_{\parallel}$ is determined by the relative alignment of the out-plane magnetic field $H_{z}$ with respect to the out-of-plane component $m_{z}^{\rm bubble}$  of the magnetization inside the bubble domain, one can replace ${\rm sgn}( V_{\parallel}(H_{z},0))$ by ${\rm sgn}(m_{z}^{\rm bubble}H_{z})$. Then, ${\rm sgn}(V_{B})$ can be rearranged as
\begin{equation}
{\rm sgn}(V_{B})={\rm sgn}(\kappa_{\rm DMI})\cdot {\rm sgn}(\theta),
\tag{S12}
\label{S12}
\end{equation}
where the DW chirality $\kappa_{\rm DMI}$  is defined by $H_{\rm DMI}$$m_{z}^{\rm bubble}$  inside the DW at the rightmost point of the bubble domain and the tilting angle $\theta$ of the magnetic field is defined by ${\rm atan}⁡(H_{x}/H_{z})$.

We confirmed the dependence on ${\rm sgn}(\theta)$ by a repeated experiment with opposite sign of $\theta$ (not shown). Such dependence on ${\rm sgn}(\theta)$ can be also verified even in the present experimental results, by rotating the observation coordinate by 180 degree with respect to the $z$ axis. On the other hand, to confirm the dependence on $\kappa_{\rm DMI}$, we repeated the experiment by use of Pt/Co/MgO films that are known to have the left-handed chirality~\cite{SE}, opposite to the right-handed chirality in the Pt/Co/Pt films~\cite{SG,HB}. Figure S1 summarizes the results from the Pt/Co/MgO films. The results truly show that the direction of the bubble motion in the Pt/Co/MgO films is opposite to that of the Pt/Co/Pt films (Fig. 2), verifying Eq. (S12).

\makeatletter 
\renewcommand{\thefigure}{S\@arabic\c@figure}
\makeatother

\begin{figure}[b]
\includegraphics[width=6.5 cm]{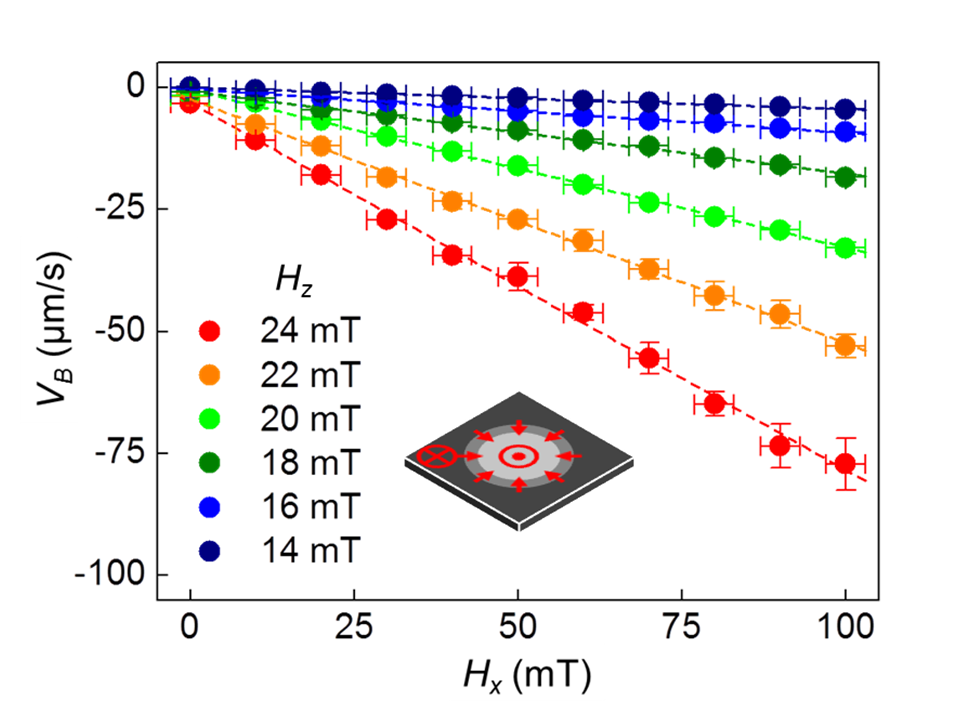}
\caption{
$V_B$ of the bubble motion in Pt/Co/MgO film, with respect to $H_x$ for several $H_z$. The inset illustrates the expected $\hat{m}_{\rm DW}$  (red arrows) with the left-handed chiral DW configuration.
The dashed lines show the best linear fit from Eq. (S9).}
\label{Fig1S}
\end{figure}